\documentclass[superscriptaddress,showpacs,preprintnumbers,amsmath,amssymb,
nofootinbib,twocolumn,aps,prd,10pt]{revtex4-1}
\usepackage{amssymb,amsmath}
\usepackage{epsfig}
\usepackage{xcolor}
\usepackage[utf8]{inputenc}
\usepackage{stmaryrd}
\usepackage{mathrsfs}
\usepackage{mathalfa}
\usepackage[normalem]{ulem}

\newcommand{\G}{\mathcal{G}}

\newcommand{\be}{\begin{equation}}
\newcommand{\ee}{\end{equation}}
\newcommand{\ba}{\begin{eqnarray}}
\newcommand{\ea}{\end{eqnarray}}
\newcommand{\beg}{\begin{gather*}}
\newcommand{\eng}{\end{gather*}}
\newcommand{\hh}{,\hspace{0.5cm}}

\newcommand{\lap}{\triangle}

\newcommand{\ind}[1]{{\mbox{\scriptsize #1}}}

\newcommand{\ts}[1]{{\boldsymbol{#1}}}

\def\XXint#1#2#3{{\setbox0=\hbox{$#1{#2#3}{\int}$ }
\vcenter{\hbox{$#2#3$ }}\kern-.6\wd0}}

\usepackage[makeroom]{cancel}
\usepackage[caption=false]{subfig}
\usepackage[colorlinks=true,
            citecolor=green,
            linkcolor=red,
            filecolor=cyan,
            urlcolor=magenta,
            backref=false]{hyperref}

\newcommand{\dd}{\mbox{d}}

\newcommand{\vf}{\langle \varphi^2(x)\rangle }

\newcommand{\rr}{$\langle \varphi^2(x)\rangle_\ind{ren}$}

\newcommand{\FT}{\langle \varphi^2(x)\rangle^{\lambda,\beta}_\ind{ren}}
\newcommand{\FTO}{\langle \varphi^2(x)\rangle^{\lambda=0,\beta}_\ind{ren}}

\begin{document}

\title{On thermal field fluctuations in ghost-free theories}

\author{Jens Boos}
\email{boos@ualberta.ca}
\affiliation{Theoretical Physics Institute, University of Alberta, Edmonton, Alberta, Canada T6G 2E1}
\author{Valeri P. Frolov}
\email{vfrolov@ualberta.ca}
\affiliation{Theoretical Physics Institute, University of Alberta, Edmonton, Alberta, Canada T6G 2E1}
\author{Andrei Zelnikov}
\email{zelnikov@ualberta.ca}
\affiliation{Theoretical Physics Institute, University of Alberta, Edmonton, Alberta, Canada T6G 2E1}

\date{\today}

\begin{abstract}
We study the response of a scalar thermal field to a $\delta$-probe in the context of non-local ghost-free theories. In these theories a non-local form factor is inserted into the kinetic part of the action which does not introduce new poles. For the case of a static $\delta$-potential we obtain an explicit expression for the thermal Hadamard function and use it for the calculation of the thermal fluctuations. We then demonstrate how the presence of non-locality modifies the amplitude of these fluctuations. Finally, we also discuss the fluctuation-dissipation theorem in the context of ghost-free quantum field theories at finite temperature.
\end{abstract}

\pacs{03.70.+k, 11.10.Lm, 11.10.Wx \hfill Alberta-Thy-03-19}

\maketitle

%%%%%%%%%%%%%%%%%%%%

\section{Introduction}
Infinite-derivative field theories emerge in various branches of physics. The subclass of these non-local field theories, which are called ghost-free theories, are of particular interest because of their nice properties.

First of all, non-local ghost-free field theories and especially ghost-free gravity have better ultraviolet behavior than their local counterparts. They have been extensively studied in a large number of publications especially in the context of the resolution of cosmological as well as black hole singularities \cite{Tomboulis:1997gg,Biswas:2011ar,Modesto:2011kw,Biswas:2013cha,Shapiro:2015uxa,Biswas:2010zk,Biswas:2013kla,Conroy:2015nva,Modesto:2017sdr,Buoninfante:2018xiw,Koshelev:2018hpt,Kilicarslan:2018unm,Kilicarslan:2019njc,Edholm:2019ypv,Frolov:2015bta,Frolov:2015usa,Frolov:2015bia,Boos:2018bxf}. Secondly, ghost-free modifications of local field theories do not lead to any new propagating degrees of freedom. As a consequence, while improving the behaviour at short distances, the predictions of ghost-free theories are very similar to those of local theories at length scales much larger than the characteristic scale of non-locality $\ell$. It would be interesting to study classical and quantum observables that distinguish ghost-free and local field theories.

One of the problems in this study is that not every method of quantization of local theories is applicable to the case of non-local theories. We have to modify and adapt traditional methods of quantization to ghost-free theories. For example, a quantization procedure based on Wick rotation from Euclidean signature to Lorentzian signature may not work for ghost-free theories. One may postulate that quantum field theory is well defined only in the Euclidean setup \cite{Biswas:2010yx,Biswas:2012ka,Buoninfante:2018mre,Buoninfante:2018gce,Asorey:2018wot,Dijkstra:2019wcz} and then try to extract information about observables in a physical domain. This approach is attractive because in the Euclidean geometry and in the local case the propagator is unique and well-defined. In the non-local case, however, the propagator picks up essential singularities in several asymptotic directions in the complex momentum plane, thereby hindering the evaluation of contour integrals by Wick rotation of correlators to the physical Lorentz signature. An alternative approach  is to define the non-local quantum theory in the physical domain with Lorentz signature without ever resorting to Wick rotation. This approach was used to study radiation effects \cite{Frolov:2016xhq} by time-dependent sources as well as quantum vacuum polarization effects and scattering on an external $\delta$-potential \cite{Boos:2018kir,Boos:2019fbu}. Technically this approach is more involved but consistent and in accordance with the physical setup of the problem.

In non-local theories some concepts like local causality and time-ordering are no longer applicable in the traditional sense. In particular, the notions of retarded and advanced propagators are to be properly generalized. In local theories the retarded (advanced) propagator $G^\ind{R(A)}(\ts{x},\ts{y})$ vanishes provided $\ts{x}$ lies everywhere outside the future (past) null cone of the point $\ts{y}$. As it has been suggested by DeWitt \cite{DeWitt:1965jb}, in non-local theories this boundary condition is to be replaced by the asymptotic condition that the retarded (advanced) propagator vanishes only in the ``remote past'' (``remote future''). Similarly, the boundary conditions for the Feynman propagator have to be replaced by suitable asymptotic conditions. This approach is well-defined for the computation of scattering amplitudes in the presence of an external potential, despite the fact that there appear some acausal effects in the vicinity of the potential.

In the present letter we study thermal fluctuations of a scalar field interacting with an external $\delta$-potential in the context of non-local, ghost-free quantum field theory. We consider the field $\hat{\varphi}$ obeying the following equation:
\begin{align}
\left[ \hat{{\cal D}} - V(x) \right] \hat{\varphi}(t,x) = 0 \, .
\end{align}
In ghost-free infinite-derivative theories the non-local operator $\hat{\cal D}$ is assumed to be of the form
\begin{align}
{\cal D}(z) = z \exp[f(z)] \, , \quad  z=\Box-m^2 \, ,
\end{align}
where the function $f(z)$ is an entire function (and therefore has no poles in the complex plane). Then, the inverse of this function
\begin{align}
{\cal D}^{-1}(z)\equiv { F(z)/ z}\hh F(z)=\exp[-f(z)]\, ,
\end{align}
has only one pole at $z=0$. The latter property implies that the propagator $\hat{\cal D}^{-1}$ does not have ghosts at tree level. We call the operator $F(\Box-m^2)$ a \emph{form factor}. When this operator acts on a function $\varphi$ that satisfies $(\Box-m^2)\varphi=0$ the form factor reduces to a constant, $F(0) = 1$.

However, there has been a significant amount of research on the higher-loop pole structure in the context of this class of theories \cite{Shapiro:2015uxa,Asorey:2018wot,Calcagni:2018gke}. The lack of new poles, at least at tree level, is an important difference of ghost-free theories to higher-derivative theories, where the propagator always picks up a non-trivial pole structure.

The simplest choice of the function is $f(z)=(-\ell^2 z)^N$, where $N$ is a positive integer number and $\ell$ is the length (or time) of non-locality. We shall refer to these non-local theories as $\mathrm{GF_N}$. It has been shown \cite{Frolov:2016xhq} that $\mathrm{GF_1}$ theory suffers from instabilities at high frequencies, while $\mathrm{GF_N}$ theories for all even N are immune to these instabilities. Note that in the case $f=0$ (and, consequently, $F=1$) one recovers the local theory.

In the absence of the potential $V$, the vacuum and thermal Hadamard functions in the non-local and ghost-free theory are the same. However, in the presence of the potential they become different. For a special case, when the potential $V$ does not depend on time and has a $\delta$-like form, both problems can be solved exactly. We shall refer to Green functions that are modified by the presence of the $\delta$ potential as \emph{dressed} Green functions.

In the present letter we focus on the study of the $\mathrm{GF_2}$ scalar theory at finite temperature. In particular, we study how non-local effects modify the quantum and thermal fluctuations of this quantum field in the presence of a static $\delta$-potential. This exactly solvable problem is a natural first step towards the understanding of the Casimir effect and other vacuum polarization effects at finite temperature in the framework of ghost-free theories.

The letter is organized as follows. After general remarks and fixation of notation in Section~II we obtain exact expressions for the dressed Hadamard functions, both in the local and ghost-free theories, in Section~III and IV, respectively. The thermal Hadamard functions, obtained in Section~IV, are used in Section~V for the calculation of the thermal fluctuations. We also study the dependence of these quantities on the temperature and the height of the potential, and compare the thermal fluctuations for local and non-local cases. Finally, Section~VI contains a summary and the application of the fluctuation-dissipation theorem \cite{Nyquist:1928zz,Callen:1951vq,Landau:1980a,Landau:1980b} to the case of non-local ghost-free field theories.

\section{Ghost-free theories with an external potential}
In this letter we will limit our discussion to a massive real scalar field $\varphi(t,x)$ in two-dimensional Minkowski spacetime in the presence of an external potential. This theory is linear, albeit at the cost of Poincar\'e invariance due to the presence of an external potential (which we shall assume to be static). Then it is possible to perform a Fourier analysis and simplify the study to that of each Fourier mode $\varphi_\omega(x)$ such that
\begin{align}
\varphi(t,x) = \int\limits_{-\infty}^\infty \frac{\dd \omega}{2\pi} e^{i\omega t} \varphi_\omega(x) \, ,
\end{align}
where due to the real-valuedness of $\varphi(t,x)$ one has
\begin{align}
\varphi_{-\omega}(x) = \varphi_\omega^\ast(x) \, .
\end{align}
The action for the Fourier modes in two dimensions is
\begin{align}
\label{eq:action}
S_\omega = \frac12 \int \dd x \, \varphi_\omega(x) \left[ \hat{\mathcal{D}}_\omega - V(x) \right] \varphi_\omega(x) \, ,
\end{align}
where $\hat{\mathcal{D}}_\omega$ denotes an operator whose precise form we shall discuss below. The equation of motion is
\begin{align}
\label{eq:eom}
\left[\hat{\mathcal{D}}_\omega - V(x) \right]\varphi_\omega(x) = 0 \, .
\end{align}
In the local case one has
\begin{align}
\hat{\mathcal{D}}_\omega^\text{loc}=\lap_{\omega} = \partial_x^2+\omega^2  - m^2 \, ,
\end{align}
that is, $\hat{\mathcal{D}}_\omega$ corresponds to the Fourier transform of the d'Alembert operator, and $m>0$ denotes the mass of the real scalar field. In the class of $\mathrm{GF_N}$ theories, the operator $\hat{\mathcal{D}}_\omega$ takes the form
\begin{align}
\hat{\mathcal{D}}_\omega^\text{GF} &= \exp\left[(-\lap_{\omega}\ell^2)^N \right] \lap_{\omega} \, ,
\end{align}
corresponding to the form factor
\begin{align}
F_\omega = \exp\left[-(-\lap_{\omega}\ell^2)^N \right] \, .
\end{align}
Here, $\ell>0$ is a new fundamental constant of the theory. It has the dimension of length (or time) and determines the scale where the modification of the theory due to its non-locality becomes important.

It is interesting that for $N=1$ this operator is reminiscent of that encountered in $p$-adic string theory \cite{Frampton:1988kr}. However, the theory \eqref{eq:action} for $N=1$ potentially leads to instabilities \cite{Frolov:2016xhq,Boos:2019fbu}. In what follows we  focus our discussion on the case $N=2$.

Let us also mention that ghost-free theories contain infinitely many derivatives, and hence they are non-local. This can be seen easily by performing a field redefinition
\begin{align}
\widetilde{\varphi}_\omega = \left(F_\omega\right)^{1/2} \varphi_\omega(x) \, ,
\end{align}
upon which the kinetic term is transformed into the standard expression
\begin{align}
\label{wv}
\widetilde{\varphi}_\omega \hat{D}_\omega^\text{loc} \widetilde{\varphi}_\omega
\end{align}
but the interaction with the external potential becomes non-local. \iffalse:
\begin{align}
\widetilde{V}(x-y)\widetilde{\varphi}_\omega(x)\widetilde{\varphi}_\omega(y) \, , \quad \widetilde{V}(x) = \left(F_\omega\right)^{-1} V(x) \, .
\end{align}\fi
In the present letter we shall not follow this approach, but rather work with the action as presented in Eq.~\eqref{eq:action}.

\section{``$\lambda$--dressing'' of Hadamard functions in the local theory}
We are interested in the calculation of the quantum average of the square of the quantum field $\hat{\varphi}$. This object can be obtained as a properly
regularized coincidence limit of the corresponding Hadamard function
\begin{align}
\label{eq:def:hdm}
G{}^{(1)}_{\bullet}(\ts{x},\ts{x'}) = \langle \hat{\varphi}(\ts{x})\hat{\varphi}(\ts{x'}) + \hat{\varphi}(\ts{x'})\hat{\varphi}(\ts{x}) \rangle_\bullet \, ,
\end{align}
where ``$\bullet$'' denotes the state over which the quantum average is to be taken.

Since we are considering only static problems, it is convenient to deal with the Fourier transform of an Hadamard function
\begin{align}
\label{eq:def:hdm-fourier}
G{}^{(1)}_{\omega,\bullet}(x,x')= \langle \hat{\varphi}_\omega(x)\hat{\varphi}_{-\omega}(x') + \hat{\varphi}_{-\omega}(x')\hat{\varphi}_\omega(x) \rangle_\bullet \, .
\end{align}
Let us denote
\begin{align}
\vf_{\omega,\bullet}={1\over 2}G{}^{(1)}_{\omega,\bullet}(x,x)\, .
\end{align}
The integration over the frequency $\omega$ of this quantity (after its proper renormalization) then gives \rr.

In the local theory the Hadamard function $G{}^{(1)}_{\omega,\bullet}(x,x')$ is
a solution of the homogeneous equation
\begin{align}
\label{eq:hdm-local-def}
\hat{\mathcal{D}}_\omega^\text{loc} G{}^{(1)}_\omega(x,x') &= 0 \, .
\end{align}
Because of Eq.~\eqref{eq:def:hdm-fourier} this solution should be symmetric with respect to the change of its arguments $x$ and $x'$. In the zero-temperature and thermal cases one has
\begin{align}
\label{hf1}
G{}^{(1)}_\omega(x,x') &= \frac{\cos[\varpi(x-x')]}{2\varpi}\theta(\varpi^2) \, , \\
\label{hf2}
G{}^{(1)}_{\omega,\beta}(x,x') &= \coth\left( \frac{\beta|\omega|}{2} \right) \frac{\cos[\varpi(x-x')]}{2\varpi}\theta(\varpi^2) \, ,
\end{align}
respectively. Here $\varpi = \sqrt{\omega^2-m^2}$ and $\beta = 1/(k_B T)$. One also has
\begin{align}
\coth\left( \frac{\beta|\omega|}{2} \right) = 1 + 2n_{|\omega|,\beta} \, , \quad
n_{\omega,\beta} = \frac{1}{e^{\beta\omega}-1} \, .
\end{align}
In the limiting case $T\rightarrow 0$, that is, $\beta \rightarrow \infty$, one finds
\begin{align}
G{}^{(1)}_{\omega,\beta\rightarrow\infty}(x,x') = G{}^{(1)}_{\omega}(x,x')
\end{align}
and one recovers the zero-temperature expression.

In the presence of the potential $V(x)=\lambda\delta(x)$, a solution of \eqref{eq:eom} can be obtained by using the Lippmann--Schwinger representation \cite{Lippmann:1950zz}
\begin{align}
\varphi_\omega(x) = \varphi^0_\omega(x) - \int\limits_{-\infty}^\infty\!\! \dd y \, G{}^\text{R}_\omega(x-y)V(y)\varphi_\omega(y) \, .
\end{align}
Here $\varphi_\omega^0(x)$ is a free solution, that is, a solution in the absence of the potential. $G{}^\text{R}_\omega(x)$ is the retarded Green function of the local equation with $V=0$,
\begin{align}
\hat{\mathcal{D}}_\omega^\text{loc} G{}^\text{R}_\omega(x) &= -\delta(x) \, ,
\end{align}
which is given by
\begin{align}
G_\omega^\ind{R}(x) = \frac{i\epsilon_\omega}{2\varpi} e^{i\epsilon_\omega\varpi |x|} \, , \quad \epsilon_\omega = \text{sign}(\omega) \, .
\end{align}
Provided $1+\lambda G^\ind{R}_\omega(0)\not=0$ one finds that
\begin{align}
\label{dress}
\varphi_\omega(x) &= \varphi^0_\omega(x)-\Lambda^\ind{loc}_\omega G{}^\text{R}_\omega(x)\varphi^0_\omega(0)\, , \\
\Lambda^\ind{loc}_\omega &= \frac{\lambda}{1+\lambda G{}^\text{R}_\omega(0)} \, .
\end{align}
The relation \eqref{dress} transforms a solution of a free equation into a solution of the equation with the $\lambda\delta(x)$ potential. We shall call this procedure $\lambda$--dressing and  the function $\varphi_\omega(x)$ a $\lambda$--dressed solution.

Similarly, one can apply a $\lambda$--dressing procedure directly to the field operators $\hat{\varphi}_{\omega}(x)$ which enter the definition of the Hadamard function \eqref{eq:def:hdm-fourier}. As a result, one finds
\begin{align}
\begin{split}
\label{eq:hdm-local}
\ts{G}^{(1)}_{\omega,\bullet}(x,x') &= G{}^{(1)}_{\omega,\bullet}(x,x') \\
&- \Lambda^\ind{loc}_\omega G_\omega^\text{R}(x) G{}^{(1)}_{-\omega,\bullet}(0,x') \\
&- \Lambda^\ind{loc}_{-\omega} G_{-\omega}^\text{R}(x') G{}^{(1)}_{\omega,\bullet}(x,0) \\
&+ \Lambda^\ind{loc}_\omega \Lambda^\ind{loc}_{-\omega} G_\omega^\text{R}(x) G_{-\omega}^\text{R}(x') G{}^{(1)}_{\omega,\bullet}(0,0) \, .
\end{split}
\end{align}
Here and in what follows we shall use the bold-face notation ``$\ts{G}$'' to refer to $\lambda$--dressed Green functions. The constructed Hadamard function has the following properties: (i) it is symmetric with respect to the change $x\to x'$, (ii) it is a solution of the homogeneous equation \eqref{eq:hdm-local-def}, and (iii) for $\lambda=0$ is coincides with the corresponding free Hadamard function. By substituting the expressions \eqref{hf1} and \eqref{hf2} into the above relation one obtains the explicit expression of the $\lambda$--dressed Hadamard functions for the vacuum and thermal states.

\section{``$\lambda$--dressing'' of Hadamard functions in the ghost-free theory}
Before presenting the results for the $\lambda-$dressed Hadamard function in the ghost-free theory it is necessary to make several remarks. Since in the absence of the potential $V$ the homogeneous solutions of the local and ghost-free equations coincide, one can easily develop the scheme of quantization in the ghost-free theory. One can simply canonically quantize the field $\widetilde{\varphi}_\omega$ defined by the equation \eqref{wv}. For such a quantization the on-shell objects of the ghost-free theory simply coincide with the corresponding objects in the local theory. In particular, a free Hadamard function defined as a solution of the homogeneous equation
\begin{align}
\label{eq:hdm-gf}
\hat{\mathcal{D}}_\omega^\text{GF} \mathcal{G}{}^{(1)}_\omega(x,x') &= 0 \, ,
\end{align}
coincides with $G{}^{(1)}_\omega(x,x')$. For the vacuum and thermal case it is given by expressions \eqref{hf1}--\eqref{hf2}.

In the presence of the potential $V$ the canonical quantization of the ghost-free theory potentially has problems connected with a proper definition of the basis of the orthonormal solutions of the field equations. For this reason we resort to the use of Green functions in order to compute the quantum and thermal fluctuations. In complete analogy to the local case we define the following expression as the $\lambda$--dressed Hadamard function in the context of ghost-free theory:
\begin{align}
\begin{split}
\label{eq:hdm-gf-dressed}
\ts{\mathcal{G}}^{(1)}_{\omega,\bullet}(x,x') &= G{}^{(1)}_{\omega,\bullet}(x,x') \\
&- \Lambda^\ind{GF}_\omega \mathcal{G}_\omega^\text{R}(x) G{}^{(1)}_{-\omega,\bullet}(0,x') \\
&- \Lambda^\ind{GF}_{-\omega} \mathcal{G}_{-\omega}^\text{R}(x') G{}^{(1)}_{\omega,\bullet}(x,0) \\
&+ \Lambda^\ind{GF}_\omega \Lambda^\ind{GF}_{-\omega} \mathcal{G}_\omega^\text{R}(x) \mathcal{G}_{-\omega}^\text{R}(x') G{}^{(1)}_{\omega,\bullet}(0,0) \, .
\end{split}
\end{align}
In the above, $\mathcal{G}_\omega^\text{R}(x)$ denotes the retarded Green function which obeys the equation
\begin{align}
\label{RR}
\hat{\mathcal{D}}_\omega^\text{GF} \mathcal{G}{}^\text{R}_\omega(x) &= -\delta(x) \, ,
\end{align}
and we defined
\begin{align}
\Lambda^\ind{GF}_\omega = \frac{\lambda}{1+\lambda \mathcal{G}_\omega^\text{R}(0)} \, .
\end{align}
The function $\ts{\mathcal{G}}^{(1)}_{\omega,\bullet}(x,x')$ is a symmetric solution of
\begin{align}
\left[\hat{\mathcal{D}}^\text{GF}_\omega-\lambda \delta(x)\right]\ts{\mathcal{G}}^{(1)}_{\omega,\bullet}(x,x')=0 \, .
\end{align}
It has the following property: In the limit when the parameter of non-locality $\ell$ tends to 0 it reproduces the corresponding local Hadamard function \eqref{eq:hdm-local}.

It should be emphasized that in order to properly define the retarded Green function in the non-local theory one should provide a proper choice of the boundary conditions for Eq.~\eqref{RR}.
DeWitt \cite{DeWitt:1965jb} has argued that in the presence of non-locality it is still possible to define asymptotic causality. Let us briefly leave Fourier space and consider two-dimensional Minkowski space, and let us denote coordinates as $\ts{x} = (x^0,x^1)$ and $\ts{y} = (y^0, y^1)$. Suppose there is a source of the field localized around $\ts{y}$, and the observer is located at $\ts{x}$. The retarded Green function in local field theory, evaluated for these two points $\ts{x}$ and $\ts{y}$, then satisfies
\begin{align}
G^\ind{R}(\ts{x},\ts{y}) \equiv 0 \quad \text{if} \quad x^0 < y^0 \, .
\end{align}
This means that the effect of the source at $\ts{y}$ on the field has to be in the causal future of this point.
In non-local quantum field theory, following DeWitt \cite{DeWitt:1965jb}, one should relax this condition and demand that the non-local retarded Green function satisfies instead
\begin{align}
\label{eq:dewitt-causality}
\mathcal{G}^\ind{R}(\ts{x},\ts{y}) \rightarrow 0 \quad \text{as} \quad x^0 - y^0 \rightarrow -\infty \, .
\end{align}

As it was demonstrated earlier \cite{Buoninfante:2018mre,Boos:2019fbu} the proper retarded Green function of the ghost-free theory, satisfying DeWitt's causality condition, is
\begin{align}
\mathcal{G}_\omega^\ind{R}(x,x') = G{}^\ind{R}_\omega(x,x') + \Delta\mathcal{G}_\omega(x,x') \, .
\end{align}
Here $\Delta\mathcal{G}_\omega(x,x') = \Delta\mathcal{G}_{-\omega}(x,x')$ is a universal non-local modification term. In the special case of $\mathrm{GF_N}$ theories it takes the form
\begin{align}
\begin{split}
\label{eq:deltaG}
\Delta\mathcal{G}_\omega(x) &= \int\limits_0^\infty \frac{\dd q}{2\pi} \cos(q x) A_\omega(q) \, ,\\
A_\omega(q) &= \frac{1-e^{-\ell^{2N}(q^2+m^2-\omega^2)^N}}{\omega^2-q^2-m^2}\, .
\end{split}
\end{align}

\begin{figure}[!htb]%
    \centering
    \includegraphics[width=0.47\textwidth]{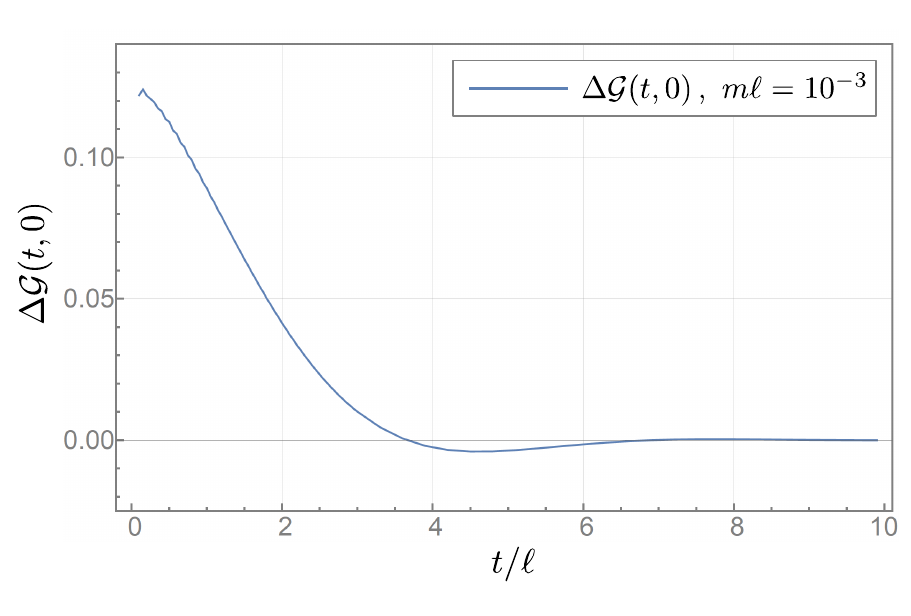}\\[-15pt]
    \caption{The non-local modification $\Delta\G(t,x)$ plotted for $\mathrm{GF_2}$ theory with a mass parameter $m\ell = 10^{-3}$. Due to Eq.~\eqref{eq:deltaG-tx-relation} it is sufficient to merely discuss, say, the timelike direction. Evidently, $\Delta\G(t,0)$ decreases rapidly in time which implies a similar property for $\Delta\G(0,x)$ in space.}
    \label{fig:deltaG-tx}
\end{figure}

At this point, let us make a remark on the real-space properties of the non-local transformation given by
\begin{align}
\begin{split}
\Delta\mathcal{G}(t,x) &= \int\limits_{\mathbb{R}^{1,1}} \frac{\dd^2 q}{(2\pi)^2} e^{iq{}_\mu x{}^\mu} A(\ts{q}^2) \, , \\
A(\ts{q}^2) &=- \frac{1-e^{-\ell^{2N}(m^2+\ts{q}^2)^N}}{m^2+\ts{q}^2} \, ,
\end{split}
\end{align}
where $\mathbb{R}^{1,1}$ denotes two-dimensional Minkowski space with the signature $\mathrm{diag}(-1,1)$ and we defined
\begin{align}
x{}^\mu = (t,x) \, , \quad
q{}_\mu = (-\omega, q) \, , \quad
\ts{q}^2 = -\omega^2 + q^2 \, .
\end{align}
We observe that the non-local modification $\Delta\G(t,x)$ is the Fourier transform of a Lorentz-invariant function $A(\ts{q}^2)$, and integrals of this type have been studied previously by DeWitt-Morette \textit{et al.} \cite{Morette:2003}. Using their results one obtains for purely timelike and spacelike distances
\begin{align}
\Delta\G(t,0) &= -\frac{1}{2\pi}\int\limits_0^\infty \dd s A(-s^2) s Y_0(s t) \nonumber \\
&\hspace{12pt} + \frac{1}{\pi^2} \int\limits_0^\infty \dd s A(s^2) s K_0(s t) \, , \\
\Delta\G(0,x) &= +\frac{1}{\pi^2} \int\limits_0^\infty \dd s A(-s^2) s K_0(s x) \nonumber \\
&\hspace{12pt} -\frac{1}{2\pi}\int\limits_0^\infty \dd s A(s^2) s Y_0(s x) \, ,
\end{align}
respectively; compare also Ref.~\cite{Buoninfante:2018mre} where similar expressions are derived. One also finds that
\begin{align}
\label{eq:deltaG-tx-relation}
\Delta\G(u,0) = -\Delta\G(0,u) \, .
\end{align}
For $t>0$ and $x>0$ and even $N$ the above integrals can be evaluated numerically in a straightforward manner since the integrands are regular functions in the entire domain of integration and they approach zero as $s \rightarrow \pm\infty$. 
See a plot of this function in the case $N=2$ for $\mathrm{GF_2}$ theory in Fig.~\ref{fig:deltaG-tx}. Evidently, the non-local modification decreases both in timelike and spacelike directions.
For odd $N$, however, this last criterion is not satisfied. These results constitute an explicit proof of the validity of DeWitt's causality condition \eqref{eq:dewitt-causality} in the context of $\mathrm{GF_2}$ theory.

For the remainder of the letter we shall return to the Fourier domain. Here, using the expression \eqref{eq:deltaG}, it is possible to show that $\Delta\G_\omega(x)$ falls off faster than any power in $1/x$ in all $\mathrm{GF_N}$ theories: Integrating by parts in Eq.~\eqref{eq:deltaG} gives
\begin{align}
\Delta\G_\omega(x)&= -\frac{1}{x} \int\limits_{-\infty}^{\infty}{\dd q\over 2\pi} \sin(q x) \frac{\dd A_\omega(q)}{\dd q} \, ,
\end{align}
where the boundary term vanishes as $q\rightarrow\infty$. These integrations by parts can be repeated ad infinitum, and, consequently, the non-local modification $\Delta\G_\omega(x)$ falls off faster than any power in $x$.

In the case of $\mathrm{GF_2}$ theory for $x=0$ the integral \eqref{eq:deltaG} can be calculated analytically and one obtains
\begin{align}
\begin{split}
\label{eq:deltaG-GF2-0}
\Delta\G_\omega(0) &= \hspace{3pt} \frac{\sqrt{2}\varpi^2\ell^3}{6\Gamma\!\left(\tfrac34\right)} {}_2F{}_2\left[ \tfrac34,\tfrac54; \tfrac32,\tfrac74; -(\varpi\ell)^4 \right] \\
&\hspace{10pt}-\frac{\Gamma\!\left(\tfrac34\right)\ell}{\pi} {}_2F{}_2\left[ \tfrac14,\tfrac34; \tfrac12,\tfrac54; -(\varpi\ell)^4 \right] \, .
\end{split}
\end{align}
To extract values of $\Delta\G_\omega(x)$ for a general point $x\not=0$, however, we have to resort to numerical calculations.

Lastly, let us remark that for even and odd $N$ the $\omega$-asymptotics are quite different. In the case $N=1$ one finds that \eqref{eq:deltaG} is exponentially divergent in $\omega$ whereas in the case $N=2$ it is a decreasing function of $\omega$ \cite{Boos:2019fbu}.

\section{Thermal fluctuations}
\begin{figure}[!htb]%
    \centering
    \includegraphics[width=0.47\textwidth]{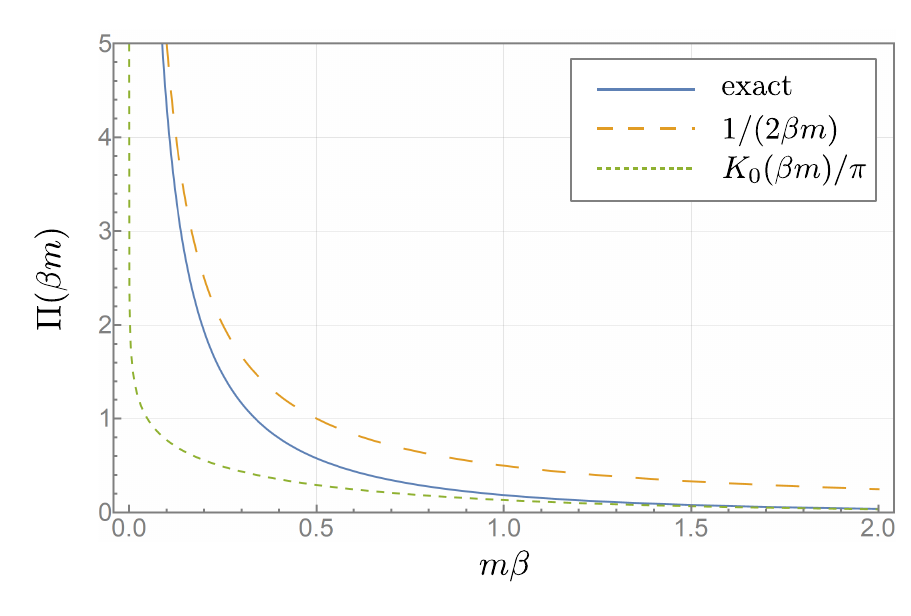}\\[-15pt]
    \caption{The thermal background $\Pi(\beta m)$ plotted as a function of the dimensionless combination $m\beta$, together with the two analytic approximations for small and large temperatures.}
    \label{fig:pp-background}
\end{figure}

Calculating the coincidence limit of the $\lambda$--dressed finite-temperature Hadamard function and subtracting the zero-point fluctuations at $\lambda=0$, $\beta=\infty$, one obtains the following expression for the renormalized finite-temperature fluctuations:
\begin{align}
\label{eq:thermal-main-result}
\FT &= \int\limits_m^\infty \frac{\dd\omega}{2\pi} \left[\ts{\mathcal{G}}^{(1)}_{\omega,\beta}(x,x) - \frac{1}{\sqrt{\omega^2-m^2}} \right] \nonumber \\
&= \Psi(x,\beta,m,\lambda,\ell) + \Pi(\beta m) \nonumber \\
\Psi(x,\beta,m,\lambda,\ell) &= \int\limits_0^\infty\frac{\dd\varpi}{4\pi} \frac{\Phi_\omega(x,\beta,m,\lambda,\ell)}{\sqrt{\varpi^2+m^2}} \nonumber \, , \\
\Phi_\omega(x,\beta,m,\lambda,\ell) &= \frac{B^2-\cos^2(\varpi x) - 2\cos(\varpi x)BC}{1+C^2} D \, , \nonumber \\
B &= 2\varpi\Delta\mathcal{G}_\omega(x) - \sin(\varpi|x|) \, , \nonumber \\
C &= 2\varpi/\widetilde{\lambda}_\omega \,  \hh \widetilde{\lambda}_\omega = \frac{\lambda}{1+\lambda\Delta\mathcal{G}_\omega(0)} \, , \nonumber \\
D &= \coth\left( \frac{\beta\sqrt{\varpi^2 + m^2}}{2} \right) \, .
\end{align}
The contribution $\Pi(\beta m)=\FTO$ is the same for both, local and non-local cases. It does not depend on $\lambda$ and describes the thermal field fluctuations in the absence of the potential. Due to translation invariance $\Pi$ does not depend on $x$ as well. Its value is given by
\begin{align}
\label{eq:pp-background}
\Pi(\beta m) = \int\limits_0^\infty \frac{\dd\varpi}{\pi\sqrt{\varpi^2+m^2}} \frac{1}{\exp(\beta\sqrt{\varpi^2+m^2}) - 1}\, .
\end{align}
The plot of this function is shown in Fig.~\ref{fig:pp-background}.
$\Pi(\beta m)$ has the asymptotics
\begin{align}
\Pi(\beta m) \approx \begin{cases} \displaystyle \frac{1}{2\beta m} & \text{ for } \beta m \ll 1 \, , \\[10pt]
\displaystyle \frac{1}{\pi} K_0(\beta m) & \text{ for } \beta m \gg 1 \, . \end{cases}
\end{align}

The function $\Psi$ depends on the height of the potential $\lambda$ and on the scale of the non-locality $\ell$. When $\ell=0$ this function describes the $x$-dependence of the thermal field fluctuations generated by the potential in the local theory. The effects of the non-locality modify the coefficients $B$ and $C$. However, these modifications are quite different. The function $\Delta\mathcal{G}_\omega(x)$ which enters the coefficient $B$ depends only on the scale of the non-locality $\ell$. At distances $x>\ell$ it becomes small. One can relate this term to the effective smearing of the $\delta$-shaped potential caused by the non-locality. 

The modification of the coefficient $C$ is of quite different nature. It can be interpreted as an effective frequency dependent change of the height of the potential $\lambda\to \widetilde{\lambda}_\omega$. This effect was discussed in our earlier paper \cite{Boos:2018kir}: the scattering amplitudes for a wave in the presence of the $\lambda\delta(x)$ potential have a specific peculiarity when its
frequency satisfies the condition
\begin{align}
1+\lambda\Delta\G_{\omega}(0) = 0 \, .
\end{align}
This frequency $\omega$ always exists provided the ``bare'' coupling $\lambda$ surpasses a critical value of
\begin{align}
\lambda \ge \lambda_\star = \frac{\Gamma\left(\tfrac14\right)}{\sqrt2\ell} \approx \frac{2.56369\dots}{\ell} \, .
\end{align}
For such a frequency the wave is totally reflected by the potential. This effect is connected with destructive interference created by the non-locality. At this frequency the coefficient $C$ vanishes, but it does not produce a significant effect on the field fluctuations. However, the effective frequency dependent modification of the height of the potential $\lambda\to \widetilde{\lambda}_\omega$ slightly modifies the amplitude of the fluctuations at distances from the potential larger that $\ell$.

\subsection{Numerical results}
Let us now resort to a numerical evaluation of the thermal fluctuations $\Psi$. Before doing so, however, let us address some physical considerations regarding the relevant choice of parameters. The function $\Psi$ depends on the parameters $x$, $\beta$, $m$, $\lambda$, and $\ell$, but there are two conditions on this parameter space that we shall demand for physical reasons.

The thermal bath at temperature $T$ contains real quanta of scalar field only when the temperature is large enough. In order to deal with this case we require
\begin{align}
\label{eq:cond1}
\frac{m}{T} = \beta m < 1 \, .
\end{align}
The effect of the external potential will be more profound if the temperature is not too high, such that the scalar field quanta will ``feel'' its presence. For this reason we assume
\begin{align}
\label{eq:cond2}
\lambda \beta > 1 \, .
\end{align}
First, we consider the local theory and insert $\Delta\G_{\omega}(x)=0$ into Eq.~\eqref{eq:thermal-main-result}. The methods for the numerical evaluation have been developed in detail in the appendix of Ref.~\cite{Boos:2019fbu}. In the present context the only difference is the additional appearance of the hyperbolic cotangent that does not affect the convergence behavior of the integral.

\begin{figure}[!htb]%
    \centering
    \includegraphics[width=0.47\textwidth]{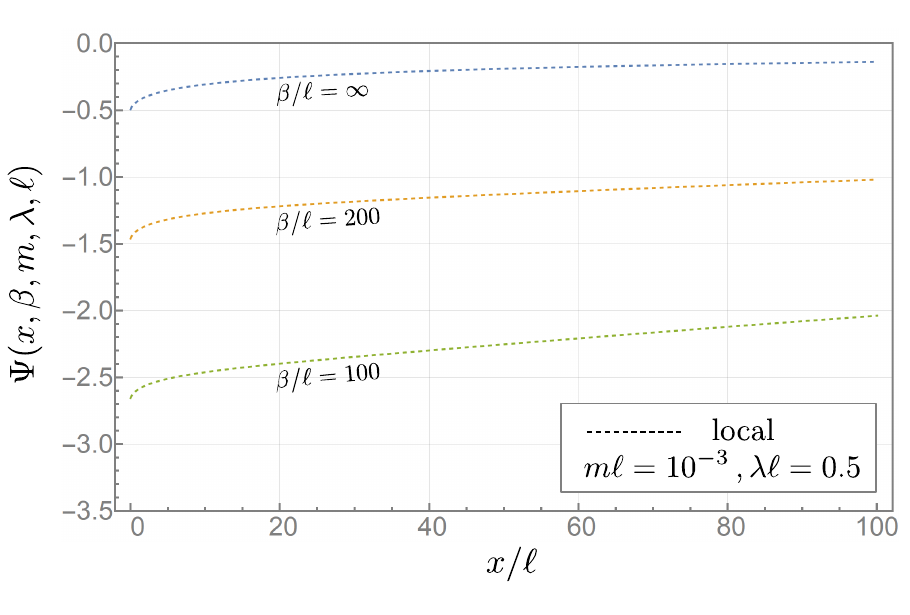}\\[-15pt]
    \caption{The local thermal fluctuations are plotted for various values of the inverse temperature, subject to conditions \eqref{eq:cond1} and \eqref{eq:cond2}, against the dimensionless distance $x/\ell$. We use the non-locality $\ell$ as a reference scale to facilitate the comparison to the non-local theory.}
    \label{fig:fluctuation-loc}
\end{figure}

Performing the numerical integration then gives the local thermal fluctuations $\Psi$ as visualized in Fig.~\ref{fig:fluctuation-loc}. In the local case, $\Psi$ depends only on the dimensionless parameters $\beta m$, $\lambda/m$, and $x m$. However, for the convenience of the comparison of these results with those of the non-local case we introduce the parameter $\ell$ in Fig.~\ref{fig:fluctuation-loc}. At the moment this parameter is arbitrary. However, when we compare these results with non-local results, we identify this parameter with the scale of non-locality.
For convenience we also plotted the zero-temperature expression $\beta=\infty$. The local thermal fluctuations exhibit a sharp peak at $x=0$ due to the presence of the potential, and they approach zero for large distance scales.

\begin{figure}[!htb]%
    \centering
    \includegraphics[width=0.47\textwidth]{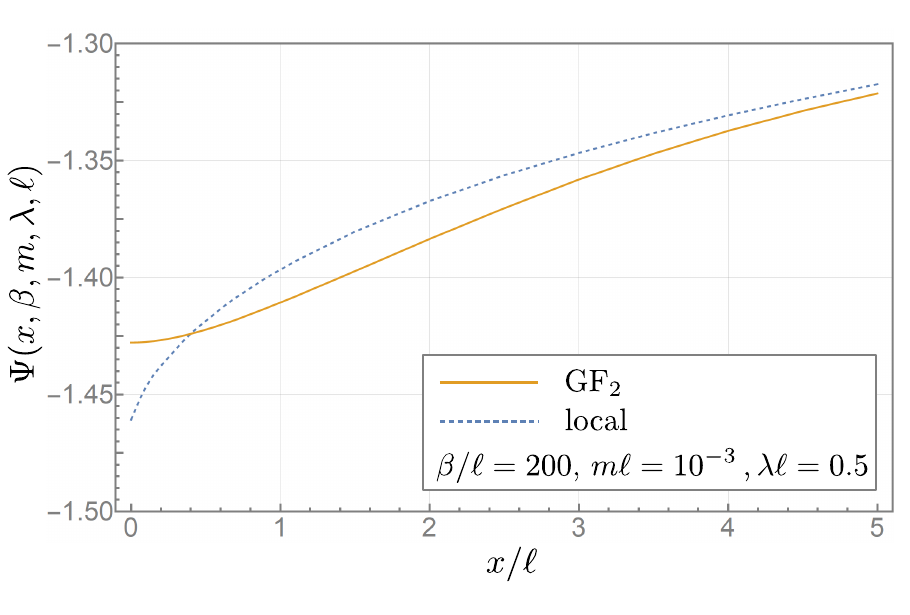}\\[-15pt]
    \caption{A comparison of the local and $\mathrm{GF_2}$ thermal fluctuations for a representative value of temperature ($\beta\ell=200$) against the dimensionless distance $x/\ell$. Mass and potential parameter are chosen in accordance to Eqs.~\eqref{eq:cond1} and \eqref{eq:cond2}.}
    \label{fig:fluctuation-gf2}
\end{figure}

It is instructive to focus on the non-locality in the region where $x/\ell \sim 1$. We expect that there will be short-scale modifications, whereas at large distance scales, $x/\ell \gg 1$, there should be no noticeable difference between the local and non-local thermal fluctuations. As we shall now see, the calculations confirm these expectations.

Let us now switch on the non-locality in Eq.~\eqref{eq:thermal-main-result} and study the modifications as compared to the local case in the context of $\mathrm{GF_2}$ theory. This corresponds to setting $N=2$ in Eqs.~\eqref{eq:deltaG} and \eqref{eq:thermal-main-result}. A numerical integration then gives rise to non-local thermal  fluctuations as depicted in Fig.~\ref{fig:fluctuation-gf2}.

For large distances, $x/\ell \gg 1$, the non-local thermal fluctuations approach the local expressions. For small distance scales, $x/\ell\sim 1$, however, there is a significant difference: the non-local thermal fluctuations approach a minimal value very smoothly. In contrast to the local theory, quantities like $\langle\varphi\partial_x\varphi\rangle$ appear to be regular at $x=0$ in $\mathrm{GF_2}$ theory.

The non-local modification in terms of the effective coupling $\widetilde{\lambda}_\omega$ affects all distance scales. Therefore, in principle, in an intermediate region $\ell \ll x \ll \infty$ it should be possible to distinguish between local and non-local theories. However, numerically we found that for choices $\lambda\ell > \lambda_\star\ell$, in agreement with conditions \eqref{eq:cond1}--\eqref{eq:cond2}, there is no notable difference. The drastic features hidden in the effective coupling $\widetilde{\lambda}_\omega$ do not seem to survive the $\omega$-integration and are apparently averaged out.

The impact of non-locality appears to improve the short-scale regularity properties of the thermal scalar fluctuations without influencing its macroscopic, long-distance behavior. In that sense this result demonstrates an explicit example of UV improvement due to the non-local, ghost-free $\mathrm{GF_2}$ theory.

\section{Discussion}
In the present letter we continued our study of the field fluctuations in ghost-free theories. In our previous work \cite{Boos:2019fbu} we discussed vacuum zero-point fluctuations in a ghost-free version of a two-dimensional, real scalar field theory. In this letter we generalize these results to the case of thermal fluctuations. In order to treat this problem we made a number of assumptions. As earlier, we considered a real scalar field in two spacetime dimensions. In order to exclude infrared divergences we assumed that the field is massive. Since we have already demonstrated that \rr\ contains divergences in $\mathrm{GF_1}$ theory we focused our study on $\mathrm{GF_2}$ theory. We explicitly demonstrated that the non-local contributions to the Green's functions in $\mathrm{GF_{2n}}$ theories are fast decreasing both in space- and time-directions, while for $\mathrm{GF_{2n+1}}$ theories this property is not valid.

For our choice of the ghost-free modification of the scalar field theory, any homogeneous (free) solution of such a theory is automatically a homogeneous solution of its ghost-free version. This happens because the corresponding on-shell form-factor is identically equal to 1. As a result, the Hadamard function in the local and non-local (ghost-free) theory calculated on-shell are the same. This is true not only for the vacuum case but for any other choice of the state: in particular, for the thermal case. This implies that the on-shell field fluctuations are the same. However, if one includes a potential term $\sim \varphi^2(x)V(x)$ in the action of the corresponding free theory the situation becomes quite different. The response of the fluctuations to the presence of the potential differs in the non-local and local case. The reason of this is simple: the potential allows one to probe off-shell properties of the theory, which are certainly different in the both cases.

In order to simplify the required calculations in the presence of the potential we chose it in the simple form $V(x)=\lambda \delta(x)$. For such a potential the problem becomes exactly solvable. Namely, if one knows a solution of the corresponding (local or non-local) equations in the absence of the potential one can
obtain an exact solution of these equations in the presence of $\lambda \delta(x)$ potential. This was achieved by proper usage of the Lippmann--Schwinger representation. We referred to this construction of an exact solution as ``$\lambda$--dressing.'' This map is a linear transformation for the field variables, however it contains a non-trivial, non-linear dependence on the parameter $\lambda$. By applying the $\lambda$--dressing to the free thermal Hadamard function we found its version in the presence of the $\delta$ potential. This allowed us to calculate the change of the thermal fluctuation caused by the presence of such a potential in the context of a ghost-free theory.

The expression describing non-local modifications of $\FT$ contains the two quantities $\Delta\G_{\omega}(x)$ and $\Delta\G_{\omega}(0)$. The first one is important in the domain very close to the position of the potential, and we have shown that it decreases faster than any power in $x$. It can be interpreted as connected with the effective smearing of the potential induced by non-locality. The other contribution, $\Delta\G_{\omega}(0)$, is related to special features of the scattering amplitudes for the non-local field. As argued above, at distances from the potential larger than the scale of non-locality $\ell$, its influence on the fluctuations can be described by an effective change of coupling constant $\lambda\to\widetilde{\lambda}_\omega$. For the thermal fluctuations, however, we verified numerically that this gives only a very small effect.

Finally, let us remark on the validity of the fluctuation-dissipation theorem \cite{Nyquist:1928zz,Callen:1951vq,Landau:1980a,Landau:1980b} for the problem under the consideration. For the local theory the fluctuation-dissipation theorem has the form \cite{Landau:1980a,Landau:1980b}
\begin{align}
\vf_\omega^{\lambda,\beta} &= \coth\left( \frac{\beta\omega}{2} \right) 2\Im \left[ \ts{{G}}_\omega^\ind{R}(x,x) \right] \, ,
\end{align}
where we defined $\vf_\omega^{\lambda,\beta} = \ts{G}^{(1)}_{\omega,\beta}(x,x)$. This corresponds to the \emph{unrenormalized} thermal fluctuations evaluated in Fourier space.

A similar relation, called generalized fluctuation-dissipation theorem, is valid when in the left-hand side instead of $\vf_\omega^{\lambda,\beta}$ one substitutes the corresponding Hadamard function and substitutes in the right-hand side the point-split version of the retarded Green's function. In the non-local ghost-free theory it is possible to prove that a similar relation is valid and one has
\begin{align}
\ts{\mathcal{G}}^{(1)}_\omega(x,x') = 2 \, \text{sgn}\left(\omega\right) \Im \left[ \ts{\mathcal{G}}^\ind{R}_\omega(x,x') \right] \, .
\end{align}
In the limit $x\rightarrow x'$ this relation gives
\begin{align}
\label{eq:fdt}
\vf_\omega^{\lambda,\beta} &= \coth\left( \frac{\beta\omega}{2} \right) 2\Im \left[ \ts{\mathcal{G}}_\omega^\ind{R}(x,x) \right] \, .
\end{align}
We expect that this relation remains valid in non-local ghost-free theories not only for $\delta$-shaped potentials, but for arbitrary forms of potentials.

\section*{Acknowledgments}
We are grateful to Anupam Mazumdar (Gr\"oningen) for insightful discussions on the subject of non-local ghost-free quantum field theory. J.B.\ is grateful for a Vanier Canada Graduate Scholarship administered by the Natural Sciences and Engineering Research Council of Canada as well as for the Golden Bell Jar Graduate Scholarship in Physics by the University of Alberta.
V.F.\ and A.Z.\ thank the Natural Sciences and Engineering Research Council of Canada and the Killam Trust for their financial support.

\bibliography{Ghost_references}{}

\end{document}